\def\r#1{\mbox{\bf r}_{#1}}
\def\rp#1{\mbox{\bf r}_{#1}^{\prime}}
\documentstyle[multicol,prl,aps,epsf]{revtex}
\begin{document}

\draft
\title{On the 'Strong-Coupling' Generalization
of the Bogoliubov Model}
\author{A.Yu. Cherny}
\address{Frank Laboratory of Neutron Physics,
Joint Institute for Nuclear Research,
141980, Dubna, Moscow region, Russia}
\author{and A.A. Shanenko}
\address{Bogoliubov Laboratory of Theoretical Physics,
Joint Institute for Nuclear Research
141980, Dubna, Moscow region, Russia}
\date{July 29, 1998}

\maketitle

\begin{abstract}
A generalized Bogoliubov model of the Bose gas in the ground state
is proposed which properly takes into account both the long--range
and short--range spatial boson correlations. It concerns equilibrium
characteristics and operates with in--medium Schr\"odinger equations
for the pair wave functions of bosons being the eigenfunctions of
the second--order reduced density matrix. The approach developed
provides reasonable results for a dilute Bose gas with arbitrary
strong interaction between particles (the 'strong--coupling' case)
and comes to the canonical Bogoliubov model in the weak--coupling
regime.
\end{abstract}
\vspace{0.5cm}

\begin{multicols}{2}

\section{Introduction}

It is well--known that the Bogoliubov model (BM) of the weakly
interacting Bose gas~\cite{Bog1} is a fundamental of the theory of
the many--boson systems. The long--range spatial correlations of
bosons are properly taken into account within this model. As to the
short--range ones, there are situations when the latter need more
accurate treatment. In particular, one can mention the troubles
appearing within BM when arbitrary strong repulsion between bosons
is expressed in a nonintegrable (singular) interparticle potential
$\Phi(r)$ behaving at small separations as $\displaystyle 1/r^m \;(m
> 3)\,.$ These troubles are commonly overcome by means of using BM
with an effective, 'dressed' interparticle potential (instead of the
'bare' one, $\Phi(r)\;$) that contains all the necessary
'information' concerning the short--range boson
correlations~\cite{Lee,Brueck,Bel,Huge}.

Although there exist sufficient amount of comprehensive ways of
constructing the effective interaction potentials (~the
pseudopotential method~\cite{Lee}, various procedures based on
summation of the ladder diagrams~\cite{Brueck,Bel,Huge}~)\footnote{
All the procedures can be reduced to the ordinary two--particle
Schr\"odinger equation with the 'bare' potential.  }, it looks
interesting and promising to realize an alternative variant of
taking account of the short--range boson correlations. We mean a
generalization of BM which operates directly with the 'bare'
potential $\Phi(r)\;$ and provides a reasonable treatment of the
short--range particle correlations side by side with the long--range
ones. A generalization of BM like this is proposed in the present
Letter.  Note that we limit ourselves to the case of the zero
temperature and consider only equilibrium characteristics such as
the pair correlation function and boson momentum distribution.

The key point of generalizing BM is based on rejecting the usual way
of dealing with the Bogoliubov model. The investigation presented
concerns the second--order reduced density matrix (2-matrix). In
particular, we operate with in--medium Schr\"odinger equations whose
solutions are the eigenfunctions of the 2--matrix, or the pair wave
functions. As an in--medium interparticle potential depends on these
functions, so the cited Schr\"odinger equations are nonlinear ones.
However, they can be linearized in the weak--coupling regime as well
as for a dilute Bose gas even with strong repulsive interaction
between bosons. The former case corresponds to the canonical
Bogoliubov model. The latter variant is related to, say, the
'strong--coupling' generalization of BM.

The present Letter is organized as follows. In the second part BM is
reconsidered in the framework of the 2--matrix. The third section
concerns the pair wave functions in the generalized BM. At last, to
show a reasonable character of the approach proposed, the
zero--density limit for the Bose gas with strong repulsion between
bosons is discussed in the fourth part of the paper.

\section{The Bogoliubov model in the light of the 2--matrix}

Let us consider a homogeneous cold many--body system of $N$~spinless
bosons with the volume~$V$ and interparticle potential $\Phi(r)\,.$
Note that absence of the spin degrees of freedom simplifies the
further reasoning without a loss of generality.

To start our investigation, let us recall the necessary definitions.
The 2--matrix for the system of interest is written as follows:
\begin{equation}
\rho_2({\bf r}_1^{\prime},{\bf r}_2^{\prime};{\bf r}_1,{\bf r}_2)=
\sum_{\nu}\,w_{\nu}\,\psi_{\nu}({\bf r}_1^{\prime},
{\bf r}_2^{\prime})\;\psi_{\nu}^*({\bf r}_1,{\bf r}_2)\; ,
\label{1}
\end{equation}
where $\psi_{\nu}({\bf r}_1,{\bf r}_2)$ are usually
called~\cite{Bog2} the pair wave functions and, physically,
$\displaystyle \sum_{\nu}\,w_{\nu}=1\;(w_{\nu}\geq 0)\,.$ The
pair wave functions for bosons are symmetric with respect to the
permutation of particles and obey the standard normalization
condition
$$
\int_V\;\int_V |\psi_{\nu}({\bf r}_1,{\bf r}_2)|^2\;
d^3 r_1\;d^3 r_2 = 1.
$$
The 2--matrix is connected with the pair correlation function
\begin{equation}
F_2({\bf r}_1,{\bf r}_2;{\bf r}_1^{\prime},{\bf r}_2^{\prime})=
\langle \psi^+({\bf r}_1) \psi^+({\bf r}_2)
    \psi ({\bf r}_2^{\prime})\psi ({\bf r}_1^{\prime})\rangle
\label{2}
\end{equation}
by the expression~\cite{Bog2}
\begin{equation}
\rho_2({\bf r}_1^{\prime},{\bf r}_2^{\prime};{\bf r}_1,{\bf r}_2)=
{F_2({\bf r}_1,{\bf r}_2;{\bf r}_1^{\prime},{\bf r}_2^{\prime})
\over N(N-1)}\;.
\label{3}
\end{equation}
Here $\psi ({\bf r}_1)$ denotes the boson field operator. Knowing
the pair correlation function, one is able to calculate all the
important thermodynamic quantities~\cite{Pines}.

The most general structure of the 2-matrix of the equilibrium
many--body system of spinless bosons is given by the following
expression~\cite{Bog3,Cherny}:
$$
F_2({\bf r}_1,{\bf r}_2;{\bf r}_1^{\prime},{\bf r}_2^{\prime})=
\sum_{\omega,{\bf q}}\,\frac{N_{\omega,q}}{V}
\varphi_{\omega,q}^{*}({\bf r})
\varphi_{\omega,q}({\bf r}^{\prime})
$$
$$
\times\exp\{i{\bf q} ({\bf R}^{\prime}-{\bf R})\}
$$
\begin{equation}
+\sum_{{\bf p},{\bf q}}\frac{N_{{\bf p},{\bf q}}}{V^2}
       \varphi_{{\bf p},{\bf q}}^{*}({\bf r})
         \varphi_{{\bf p},{\bf q}}({\bf r}^{\prime})
      \exp\{i {\bf q} ({\bf R}^{\prime}-{\bf R})\},
\label{4}
\end{equation}
where $\displaystyle {\bf r} = {\bf r}_1-{\bf r}_2,\;{\bf R}
=({\bf r}_1+{\bf r}_2)/2\;.$ The quantity $\displaystyle
\varphi_{\omega,q}({\bf r}) \cdot \exp(i{\bf q}{\bf R})/
\sqrt{V}$ denotes the wave function of the $\omega-$th bound state
of the pair of bosons with the total momentum $\hbar {\bf q}\;.$
Respectively, $\displaystyle \varphi_{{\bf p},{\bf q}}({\bf r})
\cdot \exp(i{\bf q}{\bf R})/V$ stands for the wave function
of a dissociated state of the pair of bosons with the total momentum
$\hbar {\bf q}$ and the momentum of relative motion $\hbar {\bf p}
\;.$ For the characteristics $N_{\omega,q}$ and $N_{{\bf p},
{\bf q}}$ we have: $N_{\omega,q}$ is the duplicated number
of the bound pairs of the $\omega-$th species with the total momentum
$\hbar{\bf q}\; ;\;N_{{\bf p},{\bf q}}$ is the duplicated number of
the dissociated pairs with the total momentum $\hbar {\bf q}$ and the
momentum of relative motion $\hbar {\bf p}\;.$  The wave functions
$\varphi_{\omega,q}({\bf r})$ and $\varphi_{{\bf p},{\bf q}}({\bf
r})$ obey the normalization conditions
$$
\lim\limits_{V \to \infty}\int_V |\varphi_{\omega,q}({\bf r})|^2\,
d^3r=1,
$$
\begin{equation}
\lim\limits_{V \to \infty} {1 \over V}
    \int_V|\varphi_{{\bf p},{\bf q}}({\bf r})|^2\,d^3r=1,
\label{5}
\end{equation}
and have the symmetry properties
$$
\varphi_{\omega,q}({\bf r})=\varphi_{\omega,q}(-{\bf r})\,,
$$
$$
\varphi_{{\bf p},{\bf q}}({\bf r})=\varphi_{{\bf p},{\bf q}}
(-{\bf r})=\varphi_{-{\bf p},{\bf q}}({\bf r})\,,
$$
which are a consequence of the Bose statistics. Thus, the first term
in the right--hand side of (\ref{4}) represents the sector of the
bound pairs; the second one corresponds to the dissociated states.
Remark that generally speaking, one can expect a discrete index to
appear in addition to ${\bf q}$ and ${\bf p}$ for the dissociated
states in rather complicated situations. However, this does not
concern our present consideration. So, we have restricted ourselves
to the summation over ${\bf q}$ and ${\bf p}$ in the second term of
the right--hand side of (\ref{4}).

Comprehensive analysis recently fulfilled in paper~\cite{Cherny}, has
demonstrated that, in the thermodynamic limit, the correlation
function (\ref{4}) can be rewritten as
$$
F_2({\bf r}_1,{\bf r}_2;{\bf r'}_1,{\bf r'}_2)=
n_{0}^{2}\varphi(r)\varphi(r')
+\int d^{3}p\,d^{3}q\;\frac{n_{0}}{(2\pi)^{3}}
$$
$$
\times\Biggl\{\delta\left({\bf p}
-\frac{{\bf q}}{2}\right)n\left({\bf p}+\frac{{\bf q}}{2}\right)
+\delta\left({\bf p}+\frac{{\bf q}}{2}\right)
n\left({\bf p}-\frac{{\bf q}}{2}\right)\Biggr\}
$$
\begin{equation}
\times\varphi^{*}_{{\bf p}}(\r{})\varphi_{{\bf p}}(\rp{})
\exp\{i{{\bf q}}({\bf R}^{\prime }-{\bf R})\}
+{\widetilde F}_2({\bf r}_1,{\bf r}_2;{\bf r'}_1,{\bf r'}_2),
\label{5b}
\end{equation}
where $n_0$ denotes the density of the condensed particles;
$n(p)=n({\bf p})$ stands for the distribution of the noncondensed
bosons over momenta. Note that the Bose--Einstein condensation of
particles is accompanied by the condensation of the particle pairs
and, thus, by the appearance of the $\delta-$functional terms
(the off--diagonal long--range order) in the pair distribution over
momenta $\hbar\, \bf{p}$ and $\hbar\,\bf{q}$~\cite{Yang}. The
first term in the right--hand side of (\ref{5b}) is conditioned by
presence of a macroscopic number of the pairs with $q=p=0$.
Since they include only the condensed bosons, we can call them the
condensate--condensate pairs. The second term in (\ref{5b})
corresponds to the condensate--supracondensate couples. Besides a
condensed particle, they also include a noncondensed boson. At last,
${\widetilde F}_2({\bf r}_1,{\bf r}_2; {\bf r}_1^{\prime},
{\bf r}_2^{\prime})$ is the contribution made by the
supracondensate--supracondensate dissociated states of a pair and,
maybe, by its bound states. For the wave functions of the
condensate--condensate and condensate--supracondencate couples
we have
\begin{equation}
\varphi(r)=1+\psi(r),
\;\varphi_{{\bf p}}({\bf r})=\sqrt{2}
\cos({\bf p}{\bf r})+\psi_{{\bf p}}({\bf r}) \quad (p\not=0),
\label{7}
\end{equation}
where the boundary conditions
\begin{equation}
\psi(r)\rightarrow 0 \;\;(r\rightarrow \infty), \quad \;
\psi_{{\bf p}}({\bf r}) \rightarrow 0 \;\;(r\rightarrow \infty)
\label{8}
\end{equation}
take place. At small particle separations the pair wave function
$\varphi_{{\bf p},{\bf q}}({\bf r})$ is very close to the usual
wave function of the two--body problem with the relative momemtum
$p$. Therefore, for a singular interparticle potential,
when $\Phi(r) \propto 1/r^m\;(m > 3)$ at small $r$, we have
$\varphi_{{\bf p},{\bf q}}({\bf r})\rightarrow 0$ as $r\rightarrow
0\,.$ And, hence, the relations
\begin{equation}
\psi(r=0)=-1, \quad \;\psi_{{\bf p}}({\bf r}=0)=-\sqrt{2}
\label{5c}
\end{equation}
are fulfilled.

In the case of a small depletion of the zero momentum state
(it is of interest in this Letter) we can neglect the third term in
expression (\ref{5b}):
$$
F_2({\bf r}_1,{\bf r}_2;{\bf r}_1^{\prime},{\bf r}_2^{\prime})=
n_0^2\,\varphi^*(r)\,\varphi(r^{\prime})
$$
\begin{equation}
+\frac{16 n_0}{(2\pi)^3}\int d^3p\,n(2p)
   \varphi_{{\bf p}}^*({\bf r})\varphi_{{\bf p}}({\bf r}^{\prime})
      \exp\{ i 2{\bf p}({\bf R}^{\prime}-{\bf R})\}.
\label{6}
\end{equation}
As it is known, there are two physical situations when the Bose
condensate fraction is expected to be close to 1. One of them is
related to the weak--coupling regime when a small depletion of the
zero momentum state results from a weak interaction of bosons. The
Bogoliubov model is an adequate approach of investigating this case.
Another situation occurs when we deal with a dilute Bose gas with an
arbitrary strong interaction between particles (singular potential).
Here the dilution of the system gives rise to the small depletion.
In this 'strong--coupling' regime the short--range correlations
play a significant role, which is expressed in relations (\ref{5c}).
On the contrary, the weak--coupling case is specified by the
inequalities
\begin{equation}
|\psi(r)| \ll 1, \;\quad |\psi_{{\bf p}}({\bf r})| \ll 1\;.
\label{9}
\end{equation}
In particular, the Bogoliubov model is characterized by the
choice~\cite{Cherny}
\begin{equation}
|\psi(r)| \ll 1, \;\quad \psi_{{\bf p}}({\bf r}) = 0\;.
\label{10}
\end{equation}
Expressions (\ref{6}) and (\ref{10}) allow one to obtain
$$
F_2({\bf r}_1,{\bf r}_2;{\bf r}_1,{\bf r}_2)=
n^{2}g(r)=
n_{0}^{2}\Bigl(1+\psi(r)+\psi^*(r)\Bigr)
$$
\begin{equation}
+2n_0\biggl(n-n_0+\frac{1}{(2\pi)^3}\int n(k)
                \exp(i{\bf k}{\bf r})\,d^3k\biggr),
\label{11}
\end{equation}
where $n=N/V$ and $g(r)$ is the radial distribution function.
According to the weak--coupling conditions (\ref{9}) and the
approximation adopted in (\ref{6}), it is correct to neglect the
terms of the order of $\psi(r) (n-n_0)$ and $(n-n_0)^2$ in
(\ref{11}). Besides, we may choose the wave function
of the pair ground state as a real quantity~\cite{Feyn}:
$\psi(r)=\psi^*(r)\,.$ So, expression (\ref{11}) can be rewritten
as
\begin{equation}
g(r)=1+2\psi(r)+\frac{2}{(2\pi)^3n}
\int n(k)\exp(i{\bf k}{\bf r})\,d^3k.
\label{12}
\end{equation}
Let us show that (\ref{12}) does represent the result of BM. To be
convinced of this, we need the equality
\begin{equation}
\widetilde{\psi}(k)=\int\psi(r)
         \exp(-i{\bf k}{\bf r})\,d^3r=
             \frac{1}{n_0}\langle a_{{\bf k}}\,a_{-{\bf k}}\rangle
\label{13}
\end{equation}
connecting $\psi(r)$ with the boson annihilation
operators~\cite{Cherny}. At T=0 (we deal with the zero temperature
case in the present Letter) BM yields~\cite{Bog1,Bog2} the following
relations:
\begin{equation}
\langle a_{{\bf k}}\,a_{-{\bf k}}\rangle={A_k \over 1-A_k^2},\quad
n(k)={A_k^2 \over 1-A_k^2}\; ,
\label{14}
\end{equation}
where
\begin{equation}
A_k ={1 \over n_0\widetilde{\Phi}(k)}\biggl(E(k)-{\hbar^2k^2
    \over 2m} -n_0\,\widetilde{\Phi}(k)\biggr)
\label{15}
\end{equation}
and
$$
E(k)=\sqrt{\frac{\hbar^2 k^2}{m}\,n_0\widetilde{\Phi}(k)+
\frac{\hbar^4 k^4}{4 m^2}},
$$
\begin{equation}
\widetilde{\Phi}(k)=\int\Phi(r)\exp(-i{\bf k}{\bf r})\,d^3r.
\label{16}
\end{equation}
Using (\ref{12}), (\ref{13}) and (\ref{14}) one is able to arrive
at
\begin{equation}
g(r)=1+\frac{2}{(2\pi)^3\,n}\int\,{A_k\over1-A_k}\;\exp(i
{\bf k}{\bf r})\;d^3k\;.
\label{17}
\end{equation}
This relation is exactly the result of the Bogoliubov model (see
Ref.~\cite{Bog2}).

Concluding this part of the paper, let us take notice of an
interesting equation following from (\ref{13}) -- (\ref{16}) and
being important for the reasoning of the next section. It is given
by
\begin{equation}
-\frac{\hbar^2 k^2}{m} \widetilde{\psi}(k)=
\widetilde{\Phi}(k)+2\,\widetilde{\Phi}(k)\,\Bigl(n(k)+
                             n_0\,\widetilde{\psi}(k)\Bigr)\; ,
\label{18}
\end{equation}
where $n_0\,\widetilde{\psi}(k)\;$ can be replaced by
$n\,\widetilde{\psi}(k)\;$ because we agreed to neglect the terms
of the order of $(n-n_0)\,\psi(r)\;.$ From (\ref{12}) and (\ref{18})
it follows that
\begin{equation}
\frac{\hbar^2}{m}\,\nabla^2\,\varphi(r)=\Phi(r) +
n\,\int\, \Phi(|{\bf r}-{\bf y}|) \;\Bigl( g(y)-1\Bigr)\;d^3y\;.
\label{19}
\end{equation}
This looks like the Schr\"odinger equation in the Born approximation.

\section{Pair wave functions in the generalized Bogoliubov model}

As it has been noted above, this Letter addresses the
generalization of the Bogoliubov model in such a way that
the short--range correlations should be taken into account
properly side by side with the long--range ones, a small
depletion of the zero momentum state being implied while
generalizing. So, in our further investigation it is correct
to rely on expression (\ref{6}) taken beyond the weak--coupling
regime introduced by (\ref{9}).

To employ approximation (\ref{6}) for the 2-matrix, one needs
to determine the wave functions $\varphi(r)$ and $\varphi_{{\bf p}}
({\bf r})$ beyond the weak coupling. To do this, let us
consider the in--medium two--particle problem:
\begin{equation}
H_{12}\;\psi_{\nu}({\bf r}_1,{\bf r}_2)=E_{\nu}\;
\psi_{\nu}({\bf r}_1,{\bf r}_2)\;.
\label{23}
\end{equation}
The Hamiltonian $H_{12}$ of two bosons placed into the medium of
similar bosons can be represented as
\begin{equation}
H_{12}= -\frac{\hbar^2}{2m}\nabla^2_1 - \frac{\hbar^2}{2m}\nabla^2_2
+\Phi(|{\bf r}_1-{\bf r}_2|)+U_1+U_2\;,
\label{24}
\end{equation}
where $U_i\;(i=1,2)$ stands for the energy of the interaction of the
$i-$th particle with the medium. Proceeding in the spirit of
the Thomas--Fermi approach~(for details see Ref.~\cite{shan}) and,
thus, neglecting retarding effects, one is able to approximate $U_i$
in the form
\begin{equation}
U_i=(N-2)\int\Phi(|{\bf r}_i-{\bf r}_3|)
w({\bf r}_1,{\bf r}_2,{\bf r}_3)\,d^3r_3\ \ (i=1,2),
\label{25}
\end{equation}
where $w({\bf r}_1,{\bf r}_2,{\bf r}_3)$ denotes the density of the
probability of observing the third particle at the point ${\bf r}_3$
under the condition that the first and second ones are located at
${\bf r}_1$ and ${\bf r}_2$. This quantity is connected with the
third and second reduced density matrices via the relation
\begin{equation}
w({\bf r}_1,{\bf r}_2,{\bf r}_3)=
\rho_2^{-1}({\bf r}_1,{\bf r}_2;{\bf r}_1,{\bf r}_2)\,
\rho_3({\bf r}_1,{\bf r}_2,{\bf r}_3;{\bf r}_1,{\bf r}_2,
{\bf r}_3)\,.
\label{26}
\end{equation}
Using the Kirkwood superposition approximation~\cite{Hill}
$$
V^3\;\rho_3({\bf r}_1,{\bf r}_2,{\bf r}_3;{\bf r}_1,{\bf r}_2,
{\bf r}_3)
$$
\begin{equation}
\simeq g(|{\bf r}_1-{\bf r}_2|)\,
g(|{\bf r}_1-{\bf r}_3|)\,g(|{\bf r}_2-{\bf r}_3|)\;,
\label{27}
\end{equation}
one can obtain
\begin{equation}
w({\bf r}_1,{\bf r}_2,{\bf r}_3)\simeq \frac{1}{V}\,
 g(|{\bf r}_1-{\bf r}_3|)\,g(|{\bf r}_2-{\bf r}_3|)\;.
\label{28}
\end{equation}
With (\ref{25}) and (\ref{28}) we arrive at
\begin{equation}
U_1=U_2=n\,\int\, g(|{\bf r}-{\bf y}|)\,\Phi(|{\bf r}-{\bf y}|)\;
                               g(y)\;d^3y\;.
\label{29}
\end{equation}
Thus, equation (\ref{23}) taken with the specifications (\ref{24})
and (\ref{29}) separates in the usual variables ${\bf r}={\bf r}_1-
{\bf r}_2$ and ${\bf R}=({\bf r}_1+{\bf r}_2)/2$,
$$
\psi_{\nu}({\bf r}_1,{\bf r}_2)=\varphi_{{\bf p}}({\bf r})\,
\frac{\exp(i{\bf q}{\bf R})}{\sqrt{V}}\;,
$$
and yields the following relation for the wave function
$\varphi_{{\bf p}}({\bf r})$:
$$
-\frac{\hbar^2}{m}\nabla^2\,\varphi_{{\bf p}}({\bf r})+
\Phi(r)\varphi_{{\bf p}}({\bf r})
$$
\begin{equation}
+2\varphi_{{\bf p}}({\bf r})n
\int g(|{\bf r}-{\bf y}|)\Phi(|{\bf r}-{\bf y}|)
g(y)\,d^3y=\varepsilon_p\varphi_{{\bf p}}({\bf r}).
\label{31}
\end{equation}
Let us consider equation (\ref{31}). The quantity $\varepsilon_p$
can be found taking the limit $r\rightarrow \infty\,.$
Using the asymptotic relations
$$
\lim\limits_{r\rightarrow \infty } \;g(r)=1\;,\quad
\lim\limits_{r\rightarrow \infty } \;\Phi(r)=0\;,
$$
we readily obtain at $r\rightarrow \infty$
$$
\int g(|{\bf r}-{\bf y}|)\Phi(|{\bf r}-{\bf y}|)
g(y)\,d^3y\rightarrow
\int g(|{\bf r}-{\bf y}|)\Phi(|{\bf r}-{\bf y}|)\,d^3y.
$$
Thus, we arrive at
\begin{equation}
\varepsilon_p=\frac{\hbar^2\,p^2}{m}+2n\;
\int\, g(|{\bf r}-{\bf y}|)\,\Phi(|{\bf r}-{\bf y}|)\; d^3y
\label{32}
\end{equation}
rather than $\varepsilon_p=\hbar^2\,p^2/m$ which appears within the
ordinary two--body problem. Inserting (\ref{32}) into (\ref{31}),
one is able to find
$$
\frac{\hbar^2}{m}\nabla^2\varphi_{{\bf p}}({\bf r})=
-\frac{\hbar^2 p^2}{m}+\Phi(r)\varphi_{{\bf p}}({\bf r})
$$
\begin{equation}
+2\xi_{ex}n\varphi_{{\bf p}}({\bf r})
\int g(|{\bf r}-{\bf y}|)\Phi(|{\bf r}-{\bf y}|)
\Bigl( g(y)-1\Bigr)\,d^3y,
\label{33}
\end{equation}
where the equality
$$
\int\, g(|{\bf x}-{\bf y}|)\,\Phi(|{\bf x}-{\bf y}|)\;d^3y\;=\;
\int\, g(|{\bf z}-{\bf y}|)\,\Phi(|{\bf z}-{\bf y}|)\;d^3y
$$
is used. In (\ref{33}) $\xi_{ex}$ is the correcting
factor which should be introduced to compensate oversimplification
of treating the exchange effects while deriving (\ref{31}). Indeed,
the exchange between bosons in the pair is taken into consideration:
$\varphi_{{\bf p}}({\bf r})=\varphi_{{\bf p}}(-{\bf r})=
\varphi_{-{\bf p}}({\bf r})\;.$ However,
the arguments resulting in (\ref{31}) ignore the exchange between
the particles of the pair and surrounding bosons whose influence
is considered on the mean--field level, in the spirit of the
Thomas--Fermi approach. This correcting factor may
be, in general, a function of ${\bf p}$ and some other quantities
related to the problem: $\xi_{ex}=\xi_{ex}({\bf p}, \ldots)\;.$ For
${\bf p}=0$ equation (\ref{33}) has to come to (\ref{19}) provided
relations (\ref{10}) are valid, which allows one to find in the
weak--coupling regime $\xi_{ex}({\bf p}=0)=1/2$. In this case the
approximation $\xi_{ex}=1/2$ can also be employed for ${\bf
p}\not=0$ owing to a small mean--absolute value of a boson momentum.
Moreover, we expect that in the situation of the large condensate
fraction, the choice $\xi_{ex}=1/2$ is correct beyond the weak
coupling too. The main reason for this is that the interaction
between a couple of particles and the medium is weak in both the
cases. For the potentials with a repulsive core this is due to a
small density of surrounding bosons.

Remark, that in contrast to (\ref{19}), relation (\ref{33}) is
reduced to the usual Schr\"odinger equation of the two--body problem
in the limit $n\rightarrow 0$. The same occurs for $r \rightarrow 0$
in the situation of arbitrary strong repulsion between bosons.
Indeed, in this case $\Phi(r)\rightarrow \infty \;$ at $r
\rightarrow 0\,.$ Hence, the second term in the right--hand side of
(\ref{33}) becomes much less than $\Phi(r)$ at small $r\,.$ This
leads to conditions (\ref{5c}) fulfilled at any particle density. It
is noteworthy that (\ref{33}) with $\xi_{ex}=1/2$ coincides with one
of the basic relations of the approach developed in
papers~\cite{Our}.

An important peculiarity of equation~(\ref{33}) is that it can be
used without any divergency in the integral term in the case of a
singular interparticle potential because $$g(r)\,\Phi(r)\rightarrow
0 \quad\;(r \rightarrow 0)\,.$$ Thus, (\ref{33}) reduced to
(\ref{19}) in the weak--coupling approximation, well answer our
purpose of generalizing BM.

\section{'Strong--coupling' case}

To deal with the generalization of BM based on (\ref{6}) and
(\ref{33}) with $\xi_{ex}=1/2$, we need one more equation. This
is because the number of the functions $g(r), \;n(k),\;\varphi(r),
\;\varphi_{{\bf p}}({\bf r})$ to be determined is larger than the
number of the equations at our disposal. Within the Bogoliubov
model for a weakly interacting Bose gas, the relation additional
to (\ref{12}) and (\ref{19}) at zero temperature is of the form
\begin{equation}
n^2\;\widetilde{\psi}^2(k)=n(k)\;(n(k)+1)\, ,
\label{34}
\end{equation}
(see relations (\ref{13}) -- (\ref{16})). Remark that (\ref{34})
follows from the canonical character of the well--known Bogoliubov
transformation~\cite{Bog1,Bog2}. The question now arises if one
may employ (\ref{34}) beyond the weak coupling or not. It turns out
that (\ref{34}) yields quite reasonable results even in the case
of a dilute Bose gas with strong repulsive interaction between
bosons. To be convinced of this, let us consider equations
(\ref{6}), (\ref{33}) and (\ref{34}) in the 'strong--coupling'
regime. From (\ref{34}) it follows that
$$
n(k)=\frac{1}{2}
\left(\sqrt{1+4\,n^2\,\widetilde{\psi}^2(k)}-1\right)\,.
$$
Therefore, in the limit $n \rightarrow 0$ we arrive at
\begin{equation}
\frac{n(k)}{n} =n\;\widetilde{\psi}^2_0(k)\,,
\label{35}
\end{equation}
where $\psi_0(r)$ obeys equation (\ref{33}) taken
at $n=0$ and $p=0$:
\begin{equation}
{\hbar^2 \over m} \nabla^2\Bigl(1+\psi_0(r)\Bigr)=
                           \Bigl(1+\psi_0(r)\Bigr)\,\Phi(r)\;.
\label{36}
\end{equation}
The relation (\ref{35}) suggests that all the bosons are condensed
in the zero--density limit. So, the use of (\ref{34}) does not
contradict the common expectation concerning a large condensate
fraction in a dilute Bose gas with strong repulsive interaction.
According to equation (\ref{33}), for sufficiently low values
of $p$ we have
\begin{equation}
\varphi_{{\bf p}}({\bf r})\simeq \sqrt{2}\;\varphi(r)\;
                                        \cos({\bf p}{\bf r})\,.
\label{37}
\end{equation}
Since (\ref{35}) is valid at small boson densities $n$, one can take
approximation (\ref{37}) to investigate the thermodynamics of a
dilute Bose gas. Inserting (\ref{37}) into (\ref{6}) we obtain the
following ansatz:
\begin{equation}
g(r)=\varphi^2(r)\Bigl(1 +\frac{2}{(2\pi)^3\,n}
           \int\;n(k)\exp(i{\bf k}{\bf r})\,d^3k \Bigr)\;,
\label{38}
\end{equation}
where $\varphi(r)$ is given by equation (\ref{33}) at ${\bf p}=0$
and $\xi_{ex}=1/2$:
$$
\frac{\hbar^2}{m}\nabla^2\varphi(r)=\Phi(r)\varphi(r)
$$
\begin{equation}
+n\,\varphi(r)\int g(|{\bf r}-{\bf y}|)\Phi(|{\bf r}-{\bf y}|)
\Bigl( g(y)-1\Bigr)\,d^3y.
\label{38a}
\end{equation}
This ansatz can be used for arbitrary strong repulsion between
bosons without any divergency because
$\varphi(r)\,\Phi(r)\rightarrow 0$ at $r \rightarrow 0$ while
$\Phi(r)\rightarrow \infty\,.$ It is worth noting that ansatz
(\ref{38}) is also good for a weakly interacting Bose gas. Really,
(\ref{38}) is reduced to (\ref{12}) with the assumption $|\psi(r)|
\ll 1$ and neglect of the term containing the product
$\psi(r)\,n(k)\,.$ At last, using (\ref{38}) and (\ref{38a}) and
taking the zero density limit, one can derive
\begin{equation}
\lim\limits_{n \to 0}\;g(r)=\varphi^2_0(r)\;,
\label{39}
\end{equation}
where $\varphi_0(r)=1+\psi_0(r)\,.$ Equality (\ref{39}) is the
well--known result for the pair distribution function of the Bose
gas of strongly interacting particles~\cite{Bog1}.

\section{Conclusion}

The generalization of the Bogoliubov model of the cold Bose gas has
been proposed which is based on equations (\ref{34}), (\ref{38}) and
(\ref{38a}). They come from the more complicated set of equations
(\ref{6}), (\ref{33}) with $\xi_{ex}=1/2$ and (\ref{34}) provided
the ansatz (\ref{37}) is used. The generalization properly takes
into account the short--range boson correlations side by side with
the long--range ones. The proposed approach yields reasonable
results in the weak--coupling regime as well as in the
'strong--coupling' case of a dilute Bose gas with arbitrary intense
repulsion.  The detailed analysis of the latter variant will be
fulfilled in the forthcoming paper.  As it has been noted in the
Introduction, the in--medium Schr\"odinger equations can be
linearized not only in the weak--coupling approximation. This can
also be done in the 'strong--coupling' regime. In the former case
(\ref{38a}) is reduced to equation (\ref{19}) linear in $\psi(r)\;.$
While in the latter situation (\ref{38a}) comes to an equation
linear in $\zeta(r)=\varphi(r)-\varphi_0(r)\;$ due to the obvious
inequality $|\zeta(r)| \ll|\varphi_0(r)| (n\rightarrow 0)\;.$


\end{multicols}

\end{document}